\documentclass[fleqn,10pt]{wlscirep}
\usepackage{amsfonts,amssymb,graphics,graphicx,bbm}

\usepackage{subfigure}
\usepackage{bm}
\usepackage{color}
\usepackage{epstopdf}
\usepackage{amssymb}
\usepackage{amstext}
\usepackage{latexsym}
\usepackage{hyperref}
\usepackage{amsfonts}
\usepackage{psfrag}
\usepackage{graphicx}

\expandafter\let\csname equation*\endcsname\relax
\expandafter\let\csname endequation*\endcsname\relax
\usepackage{amsmath}

\newcommand{\ket}[1]{\vert #1 \rangle}
\newcommand{\bra}[1]{\langle #1 \vert}

\newcommand{\braket}[2]{\langle #1 \vert #2 \rangle}

\newcommand{\abs}[1]{| #1 |}
\newcommand{\eg}{{\it{e.g.}}}
\newcommand{\ie}{{\it{i.e.}}}

\begin{document}

\title{Experimental semi-autonomous eigensolver using reinforcement learning}

\author[1]{C.-Y. Pan}
\author[1]{M. Hao}
\author[1]{N. Barraza}
\author[1,2,3,4,*]{E. Solano}
\author[1,*]{F. Albarr\'an-Arriagada}

\affil[1]{International Center in Quantum Artificial Intelligence for Science and Technology (QuArtist)  and Physics Department, Shanghai University, 200444 Shanghai, China}
\affil[2]{Department of Physical Chemistry, University of the Basque Country UPV/EHU, Apartado 644, 48080 Bilbao, Spain}
\affil[3]{IKERBASQUE, Basque Foundation for Science, Plaza Euskadi 5, 48009 Bilbao, Spain}
\affil[4]{Kipu Quantum, Kurwenalstrasse 1, 80804 Munich, Germany}

\affil[*]{enr.solano@gmail.com}
\affil[*]{pancho.albarran@gmail.com}

\begin{abstract}
The characterization of observables, expressed via Hermitian operators, is a crucial task in quantum mechanics. For this reason, an eigensolver is a fundamental algorithm for any quantum technology. In this work, we implement a semi-autonomous algorithm to obtain an approximation of the eigenvectors of an arbitrary Hermitian operator using the IBM quantum computer. To this end, we only use single-shot measurements and pseudo-random changes handled by a feedback loop, reducing the number of measures in the system. Due to the classical feedback loop, this algorithm can be cast into the reinforcement learning paradigm. Using this algorithm, for a single-qubit observable, we obtain both eigenvectors with fidelities over $0.97$ with around $200$ single-shot measurements. For two-qubits observables, we get fidelities over $0.91$ with around $1500$ single-shot measurements for the four eigenvectors, which is a comparatively low resource demand, suitable for current devices. This work is useful to the development of quantum devices able to decide with partial information, which helps to implement future technologies in quantum artificial intelligence. 		
\end{abstract}

\maketitle	

\section{Introduction}
Increasing the computational capabilities of machines is an essential field in artificial intelligence. In this context, machine learning algorithms have emerged with great force in the last decades~\cite{Russell1995Book, Metha2019}. This class of algorithms can be divided into two families, learning from big data and learning from interactions. Learning from big data can be classified into two categories, supervised and unsupervised learning. In the supervised learning paradigm, we have a set of labeled data named training data, from which we want to infer some classification function to sort unlabeled new data. Unsupervised learning algorithms do not use training data. In this paradigm, the goal is to extract the statistical structure of an unsorted data set and divide it into different groups according to some criteria (clustering problem)~\cite{Ghahramani2003, Kotsiantis2007, Wiebe2012, Lloyd2013, Rebentrost2014, Zhaokai2015}.

In the category of learning from interactions we have the Reinforcement Learning (RL) algorithms~\cite{Sutton2018Book,Lamata2017,Jaderberg2019,Kaelbling1996,dong2008quantum,Mnih2015,Riedmiller2009,Yu2019,F2018,Littman2015}. The idea in this paradigm is that a known and manipulable system called \textit{agent} ($A$) interacts with a non-manipulable system called \textit{environment} ($E$). Here, the goal is to optimize a task $\mathcal{G}(A, E)$, which depends on the state of $A$ and $E$. For this, we use feedback loops to change the state of $ A $ using the information extracted from the interaction with $ E $. Some impressive and recent examples of RL are the AI players for different strategy games like Go~\cite{Silver2017}, Chess~\cite{Silver2018}, or StarCraft II~\cite{Vinyals2019}. 

On the other hand, it has been shown that quantum computing~\cite{Nielsen2010Book} can overcome some fundamental limits of classical computing, \eg, in searching problems~\cite{Grover1996}, factorization algorithms~\cite{Shor1997}, solving linear equation systems~\cite{Harrow2009, Cai2013}, and for linear differential equations~\cite{Xin2020}. Therefore, it was natural to merge machine learning techniques with the advantages of quantum computing in the topic known as Quantum Machine Learning (QML)~\cite{Biamonte2017Review, Schuld2015Review, Gao2018, Dunjko2016, Schuld2019, Lau2017, Wittek2014Book, Lamata2020}. 

With the development of Noisy Intermediate-Scale Quantum (NISQ) devices~\cite{Preskill2018}, the research on simple quantum information protocol (suitable for NISQ quantum computers) and the research in QML has grown in the last years. The IBM quantum computer is one of the most famous open NISQ devices, which can be programmed using Qiskit~\cite{Qiskit}, an open-source python package, to create and run quantum programs using the IBM quantum cloud service~\cite{IBMQ}.

One of the most useful algorithms for linear algebra, and hence for quantum mechanics, are the quantum eigensolvers. The hybrid quantum-classical algorithms like variational quantum eigensolver (VQE)~\cite{Peruzzo2014, McClean2016, Kandala2017} take advantage due to its easy implementation in NISQ devices. The main idea of this class of algorithm is to calculate some expectation value (like energy) with a quantum processor, and then use a classical optimizer (like variational one) to reach the solution~\cite{Lavrijsen2020}. Nevertheless, it has been recently proposed an algorithm that uses a quantum optimizer~\cite{Wei2020}. Each iteration of the classical optimizer algorithm involves many single-shot measurements in the quantum system, which are required to calculate an expectation value. The development of an algorithm with more quantum features will involve the use of a more primitive classical subroutine.

In this paper, we implement the semi-autonomous eigensolver proposed in Ref.~\cite{AlbarranArriagada2020}. The protocol can obtain an approximation of all eigenvectors for an arbitrary observable using single-shot measurements instead of expectation values. Here, we use the most basic classical subroutine, which involves only pseudo-random changes handled by the outcome of the single-shot measurement and a feedback loop. Due to this feedback loop, this algorithm can be classified in the RL paradigm. Using our protocol, we can obtain a high fidelity approximation for all eigenvectors. In the single-qubit case, we get fidelities larger than $0.97$ and larger than $0.91$ for a two-qubit observable in around 200 and 5000 single-shot measurements, respectively. This work opens the door to explore alternative paradigms in hybrid classical-quantum algorithms, which is useful for developing semi-autonomous quantum devices that decide with incomplete information.

\section{Methods}
\label{Section2}
\subsection{Basics on RL paradigm}
We briefly describe the basic components of the RL paradigm. As mentioned above, in an RL algorithm, we define two systems: the agent $ A $ and the environment $ E $. The interaction among these systems can be divided in three basic steps, the \textit{policy}, the \textit{reward function} (RF) and the \textit{value function} (VF). The policy refers to the general rules of the algorithm and can be subdivided into three stages: first, the interaction, where we specify how $A$ and $E$ interact; second, the action, which refers to how $A$ changes its perception of $E$ modifying some internal parameters; and third, the information extraction, that defines the process used by $A$ to infer information from $E$. The information extraction can be done directly by $ A $ or using an auxiliary system, named \textit{register}, if $ A $ cannot read the response of the environment. 

The RF is the criterion to reward or punish $ A $ in each iteration using the information collected from $ E $. This step is the most important in any RL algorithm because the right choice of the RF ensures the optimization of the desired task $\mathcal{G}(A, E)$. Finally, the VF evaluates a figure of merit related to the task $\mathcal{G}(A, E)$, which provides us the utility of the algorithm. The main difference between RF and VF is that the first evaluates each iteration to increase the performance locally in time without considering the history of the algorithm. At the same time, VF depends on the history of the algorithm, which takes into consideration a large number of iterations given the global performance of the algorithm.

\subsection{RL protocol}

We define the basic parts of our protocol as an RL algorithm. The state of the agent is denoted by
\begin{equation}
\ket{\mathcal{A}_k^{(j)}}=\hat{D}_k\ket{j},
\label{Eq01}
\end{equation}
where $\hat{D}_k$ is a unitary transformation to prepare the desired agent state, the state $\ket{j}$ is the initial state provided by the quantum processor in the computational basis, and the subindex $k$ denotes the iteration of the algorithm. The environment is expressed as an unknown Hermitian operator $\hat{\mathcal{O}}$ written as
\begin{equation}
\hat{\mathcal{O}}=\sum_{j} \alpha^{(j)}\ket{\mathcal{E}^{(j)}}\bra{\mathcal{E}^{(j)}},
\label{Eq02}
\end{equation}
with $\alpha^{(j)}$ and $\ket{\mathcal{E}^{(j)}}$ the $j$th eigenvalue and eigenvector of $\hat{\mathcal{O}}$, respectively. The task $\mathcal{G}$ is set to maximize the fidelity between the state of the agent, $\ket{\mathcal{A}_N^{(j)}}$, after $N$ iterations, and the eigenvectors $\ket{\mathcal{E}^{(j)}}$, or in other words, we want to find the matrix $\hat{D}_k$ that diagonalizes the observable $\hat{\mathcal{O}}$.

Now, the policy is as follows:

\textit{Interaction:} The observable $\hat{\mathcal{O}}$ generates an evolution given by the unitary transformation
\begin{equation}
\hat{E}=e^{-i\hat{\mathcal{O}}\tau}=\sum_je^{-i\alpha^{(j)}\tau}\ket{\mathcal{E}^{(j)}}\bra{\mathcal{E}^{(j)}},
\label{Eq03}
\end{equation}
where $\tau$ is a constant related with the elapsed time of the interaction. The agent state after this evolution is
\begin{equation}
\hat{E}\ket{\mathcal{A}_k^{(j)}}=\ket{\bar{\mathcal{A}}_k^{(j)}}=\sum_{\ell}c^{(\ell)}\ket{\mathcal{A}_k^{(\ell)}}.
\label{Eq04}
\end{equation}

\textit{Information extraction:} We measure the state $\ket{\bar{\mathcal{A}}_k^{(j)}}$ in the basis $\{\ket{\mathcal{A}_k^{(\ell)}}\}$. For this purpose we apply the transformation $\hat{D}^{\dagger}_k$ obtaining
\begin{equation}
\hat{D}^{\dagger}_k\ket{\bar{\mathcal{A}}_k^{(j)}}=\sum_{\ell}c^{(\ell)}\ket{\ell},
\label{Eq05}
\end{equation}
followed by a single-shot measurement in the computational basis $\{\ket{\ell}\}$ obtaining the outcome value $m$ with probability $|c^{(m)}|^2$. This outcome refers to the resulting state $\ket{\mathcal{A}_k^{(m)}}$ after the measuring process.

\textit{Action:} According to Eq. (\ref{Eq03}) if $\ket{\mathcal{A}_k^{(j)}}$ is equal to some eigenvector of $\hat{\mathcal{O}}$, we obtain $c^{(j)}=1$ in Eq. (\ref{Eq04}). Using this condition we define the next rule for the action. If the outcome is $m\ne j\Rightarrow c^{(j)}\ne 1$, then $\ket{\mathcal{A}_k^{(j)}}$ is not an eigenvector of $\hat{\mathcal{O}}$. In this case ($m\ne j$), we modify the agent for the next iteration defining operator $\hat{D}_{k+1}$ as
\begin{equation}
\hat{D}_{k+1}=\hat{D}_{k}\hat{u}_{j,m}(\theta,\phi,\lambda),
\label{Eq06}
\end{equation}
with
\begin{equation}
\hat{u}_{j,m}(\theta,\phi,\lambda)=e^{-i\lambda\hat{S}_{j,m}^{(z)}}e^{-i\theta\hat{S}_{j,m}^{(y)}}e^{-i\phi\hat{S}_{j,m}^{(z)}} ,
\label{Eq07}
\end{equation}
where,
\begin{eqnarray}
S_{j,m}^{(z)}=\frac{1}{2}\left(\ket{j}\bra{j}-\ket{m}\bra{m}\right),\nonumber\\
S_{j,m}^{(y)}=-\frac{i}{2}\left(\ket{j}\bra{m}-\ket{m}\bra{j}\right) .
\label{Eq08}
\end{eqnarray}
Then,
\begin{eqnarray}
\hat{u}_{j,m}(\theta,\phi,\lambda)=&&\cos\left(\frac{\theta}{2}\right)\left(\ket{j}\bra{j}+e^{i(\lambda+\phi)}\ket{m}\bra{m}\right)\nonumber\\
&&+\sin\left(\frac{\theta}{2}\right)\left(-e^{i\phi}\ket{j}\bra{m}+e^{i\lambda}\ket{m}\bra{j}\right)
\label{Eq09}
\end{eqnarray}
up to a global phase. Therefore, $\hat{u}(\theta, \phi, \lambda)$ is a general rotation in the $\{\ket{j},\ket{m}\}$ subspace. The angles are random numbers given by
\begin{equation}
\{\theta,\lambda,\phi\}\in w_k\cdot[-\pi,\pi],
\label{Eq10}
\end{equation}
where the range amplitude $w_k$ will be updated in each iteration according to the RF, which will be specified later. Now, for the case $m=j$, the state $\ket{\mathcal{A}_k^{(j)}}$ could be an eigenvector of $\hat{\mathcal{O}}$, then we define
\begin{equation}
	\hat{D}_{k+1}=\hat{D}_k.
	\label{Eq11}
\end{equation}
We can summarize Eqs. (\ref{Eq06}) and (\ref{Eq11}) as
\begin{equation}
	\hat{D}_{k+1}=\hat{D}_k\left[\sum_{l\ne j}\hat{u}_{l,m}(\theta,\phi,\lambda)\cdot\delta_{l,m} + \mathbb{I}\cdot\delta_{j,m}\right].
	\label{Eq12}
\end{equation}

Now, we define the reward function as
\begin{equation}
w_{k+1}=w_k\left[p\cdot\sum_{l\ne j}\delta_{l,m} + r\cdot\delta_{j,m}\right]
\label{Eq13}
\end{equation}
where $p>1$ is the punishment ratio, and $0<r<1$ is the reward ratio. This means that each time we obtain the outcome $m\ne j$, we increase the amplitude range $w_{k+1}$, because $m\ne j$ means that we are further away from an eigenvector and greater corrections are required. In the other case, when $m=j$ means that we are closer to an eigenvector, then, we reduce the value of $w_{k+1}$ obtaining smaller changes for future iterations.

Finally, the value function will be the last value of the range amplitude $w_N$ after $N$ iterations. If $w_N\rightarrow 0$ signifies that we have measured $m=j$ several times, then $c^{(j)}\approx 1$, which implies that we obtain a good approximation of an eigenvector.

\begin{figure}[t]
\centering
\includegraphics[width=0.4\linewidth]{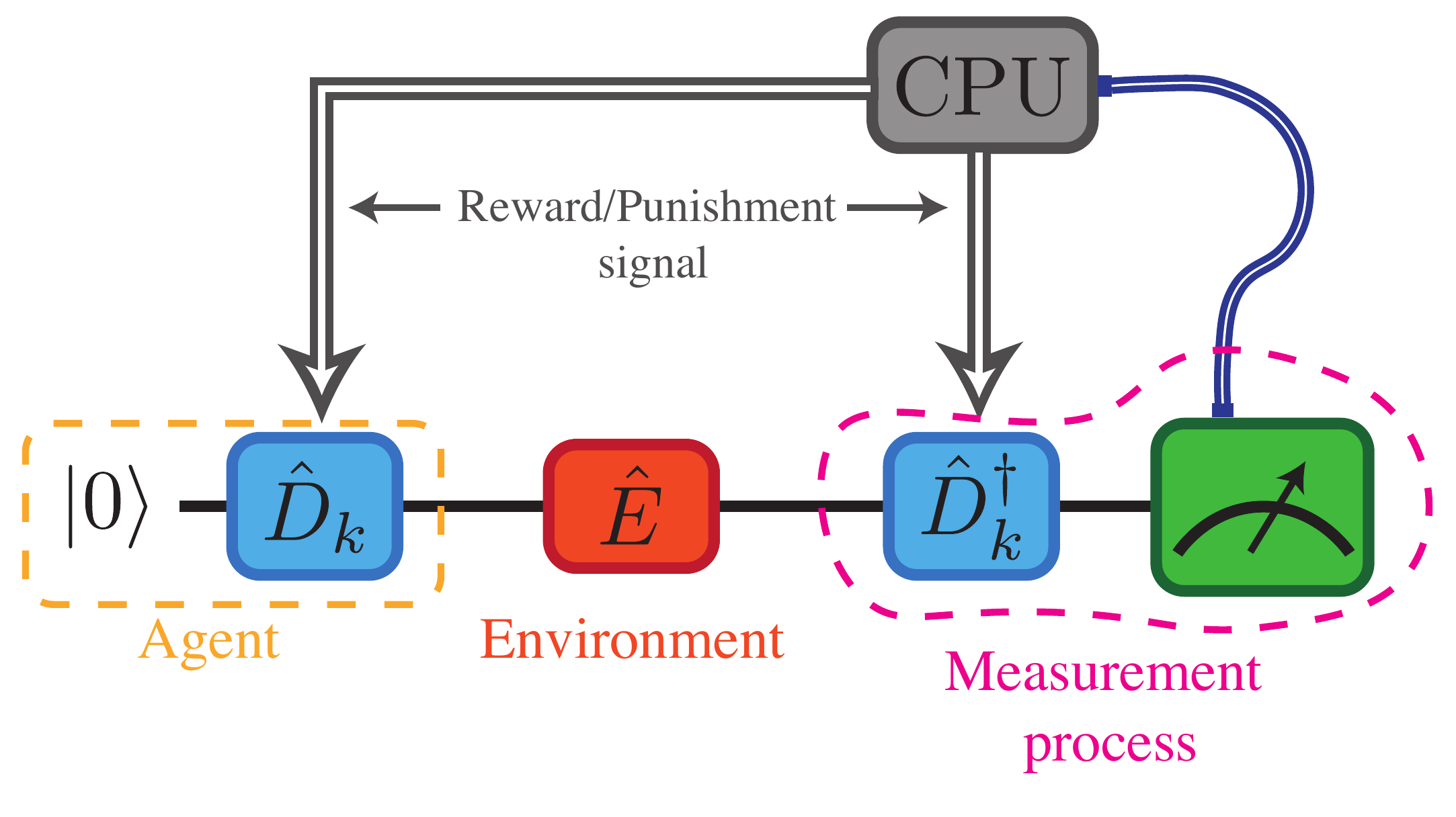}
\caption{Diagram of the single-qubit protocol. The subindex $k$ refers to the $k$th iteration. Blue lines represent the classical communication to the central processing unit. The gray arrows show feedback loops, where $\hat{D}_k$ and $\hat{D}_k^{\dagger}$ are updated according to the measurement outcome.}
\label{Fig01}
\end{figure}

\section{Results}
\subsection{Single-qubit case}
We implement the algorithm described above in the IBM quantum computer. We start with the simplest case, which is to find the eigenvectors of a single-qubit observable. Since there are only two eigenvectors, we only need to obtain one of them, because the orthogonality property can determine the second one. Figure \ref{Fig01} shows the circuit diagram for this case. As we can see in figure \ref{Fig01} the agent in each iteration is given by
\begin{equation}
\ket{\mathcal{A}_k^{(0)}}=\hat{D}_k\ket{0}.
\label{Eq14}
\end{equation}

\begin{figure}[h]
\centering
\includegraphics[width=0.4\linewidth]{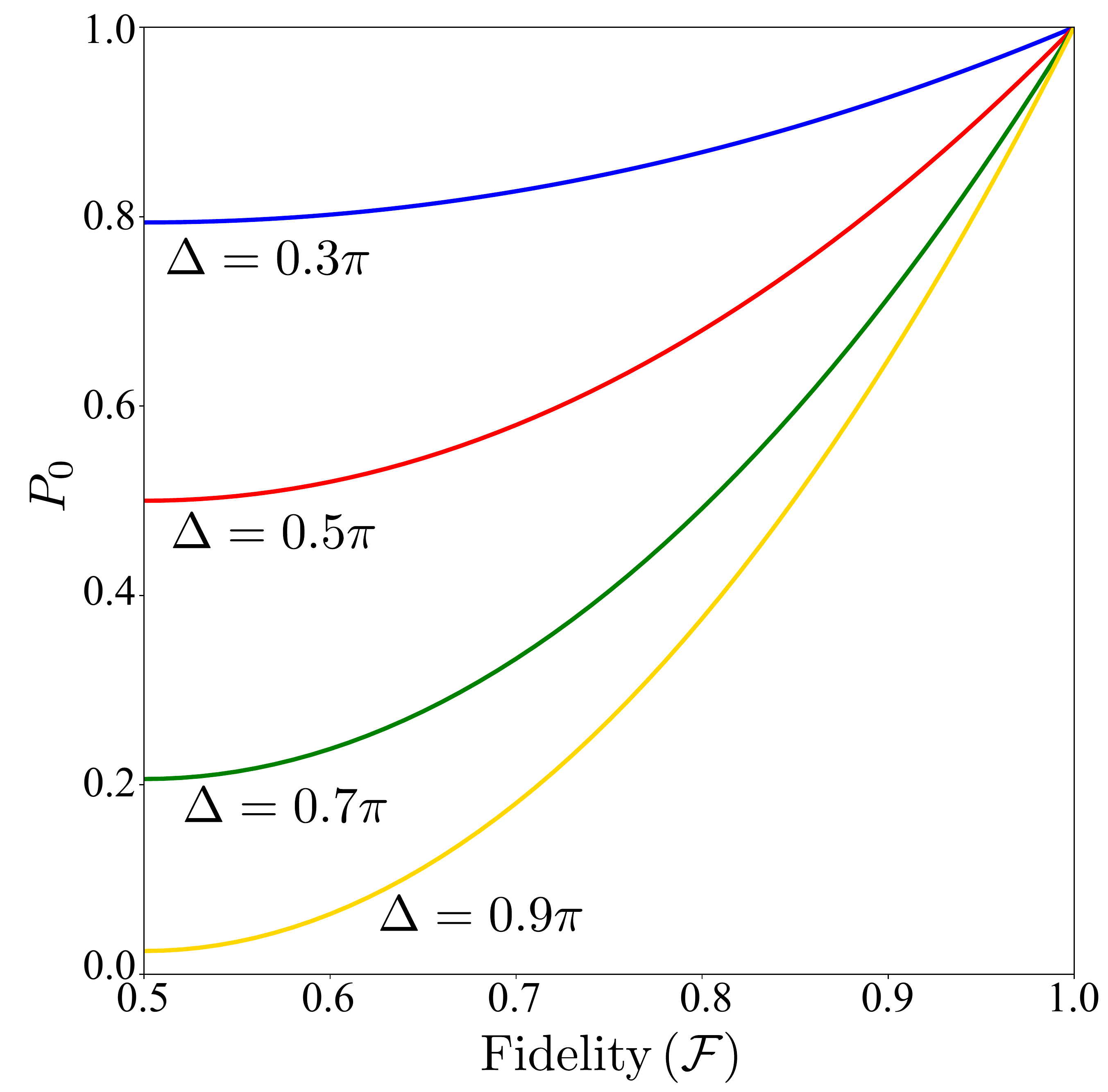}
\caption{$P_0$ as a function of $\mathcal{F}$ for different values of $\Delta$.}
\label{Fig02}
\end{figure}

In this case, we have only one the rotation ($\hat{u}_{1,0}$) of the form of Eq. (\ref{Eq07}), then, for simplicity, we redefine the operator $\hat{D}_k=\hat{D}(\theta_k,\phi_k,\lambda_k)$ as
\begin{equation}
\hat{D}(\theta_k,\phi_k,\lambda_k)=e^{-i\frac{\lambda_k}{2}\hat{\sigma}^{(z)}}e^{-i\frac{\theta_k}{2}\hat{\sigma}^{(y)}}e^{-i\frac{\phi_k}{2}\hat{\sigma}^{(z)}},
\label{Eq15}
\end{equation}
where $\hat{\sigma}^{(a)}$ is the $a$-Pauli matrix and
\begin{eqnarray}
	\theta_{k+1}=\theta_k+\Delta_{\theta}\cdot\delta_{1,m}, \nonumber\\
	\phi_{k+1}=\phi_k+\Delta_{\phi}\cdot\delta_{1,m}, \nonumber\\
	\lambda_{k+1}=\lambda_k+\Delta_{\lambda}\cdot\delta_{1,m},
	\label{Eq16}
\end{eqnarray}
with $\{\Delta_{\theta},\Delta_{\phi},\Delta_{\lambda}\}\in w_k[-\pi,\pi]$ and $w_k$ given by Eq. (\ref{Eq13}), considering only two outcomes ($m\in\{0,1\}$) and $j=0$ for the whole algorithm. The gate in Eq. (\ref{Eq15}) has the form of the general qubit-rotation provided by qiskit, therefore, it can be efficiently implemented in the IBM quantum computer. We denote by, $\mathcal{F}$, the maximum fidelity between the agent state, $\ket{\mathcal{A}_N^{(0)}}$, and one of the eigenvectors at the end of the algorithm. We find that $\mathcal{F}$ is related to the probability of obtaining the outcome $m = 0$ ($P_0$) by (see appendix $A$)
\begin{eqnarray}
	&&P_0=\frac{1-\cos(\Delta)}{2}\left[(2\mathcal{F}-1)^2-1\right] + 1\nonumber\\
	\Rightarrow&&\mathcal{F}=\frac{1}{2}\left(1+\sqrt{\frac{2(P_0-1)}{1-\cos{\Delta}}+1}\right),
	\label{Eq17}
\end{eqnarray}
where $\Delta=\tau\abs{\alpha^{(0)}-\alpha^{(1)}}$ is the gap between the eigenvalues of $\tau\hat{\mathcal{O}}$ (see Eqs. (\ref{Eq02}) and (\ref{Eq03})). Figure \ref{Fig02} shows $P_0$ as a function of the fidelity $\mathcal{F}$ for different values of $\Delta$.

\begin{figure}[h]
\centering
\includegraphics[width=0.6\linewidth]{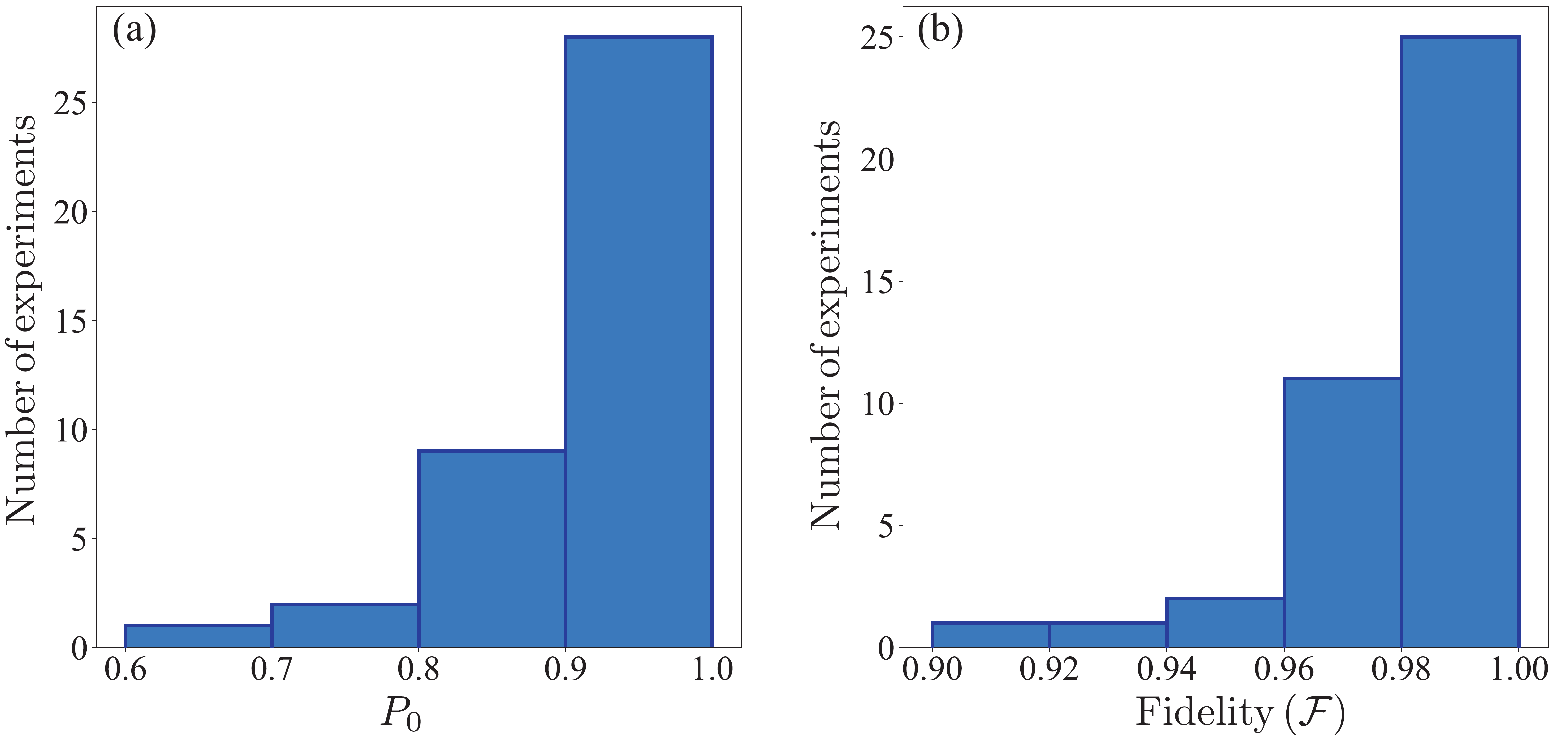}
\caption{Histograms for the results of $40$ independent experiments. with $\tau\hat{\mathcal{O}}=\frac{\pi}{2}\sigma_x$, $r=0.9$ and $p=1/r$. (a): Histogram for the probability to obtain $m=0$. (b): Histogram for the fidelity between the agent and the nearest eigenvector using Eq.(\ref{Eq17}).}
\label{Fig03}
\end{figure}

\begin{figure}[h]
\centering
\includegraphics[width=0.6\linewidth]{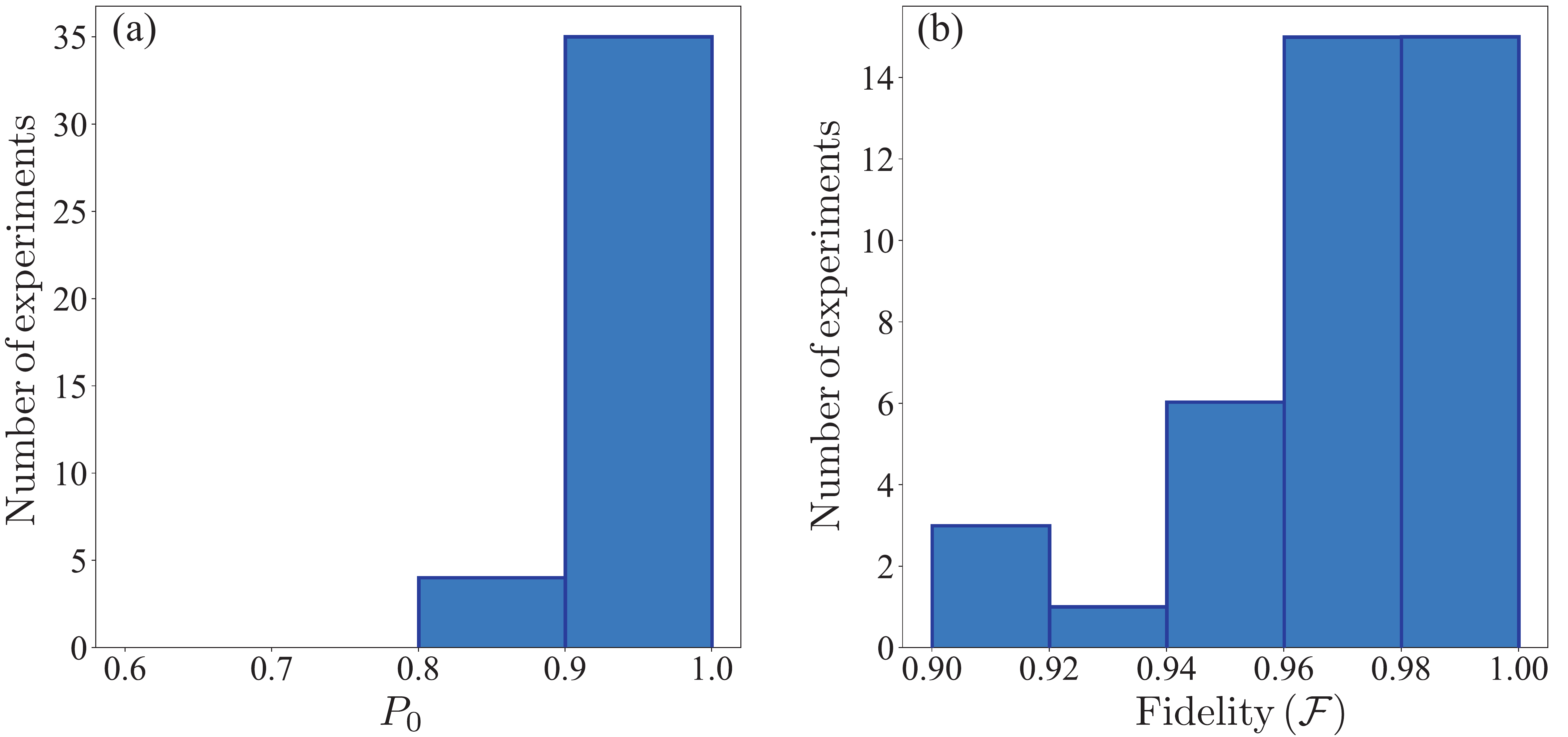}
\caption{Histograms for the results of $40$ independent experiments. with $\tau\hat{\mathcal{O}}=\frac{\pi}{4}\sigma_x$, $r=0.9$ and $p=1.5/r$. (a): Histogram for the probability to obtain $m=0$. (b): Histogram for the fidelity between the agent and the nearest eigenvector using Eq. (\ref{Eq17}).} 
\label{Fig04}
\end{figure}

For the implementation we use the initial values $\theta_1=\phi_1=\lambda_1=0$, $w_1=1$ and the quantum processor ``ibmqx2". The algorithm is run until $w_N<0.1$. Since the algorithm converges stochastically to the eigenvectors, we perform $40$ experiments in order to characterize the performance of the algorithm by the central values of the data set. Also, we compare the performances of our algorithms with the $VQE$ algorithm for the same environments using the same quantum processor. To test the algorithm, we use three different environment Hermitian operators:
\begin{equation}
1.\quad\tau\hat{\mathcal{O}}=\frac{\pi}{2}\sigma_x\Rightarrow\Delta=\pi\Rightarrow\mathcal{F}=\frac{1}{2}(1+\sqrt{P_0}).\nonumber
\end{equation}
Here, we choose the reward ratio $r=0.9$ and the punishment ratio $p=1/r$. The results of the $40$ experiments are collected in the Apendix Table 1 (Supplemental material) and summarized in the histograms of Fig. \ref{Fig03}. From figure \ref{Fig03} (a), we can see that the probability $P_0$ is bigger than 0.85 in $36$ cases, which implies, as is shown in Fig. \ref{Fig03} (b), that most cases give fidelities larger than $0.94$. Also, we have $36$ experiments with $\mathcal{F}> 0.96$, the average fidelity is $\bar{\mathcal{F}}=0.98$ and the standard deviation is $\sigma=0.019$ which represent the $2\%$ of the average fidelity $\bar{\mathcal{F}}$. Also, the average number of iterations of the algorithm in the $40$ experiments is $\bar{N}=103$, the minimum number of iterations $N_{min}=25$, and the maximum number of iterations $N_{max}=528$. This number may look large, but we remark that we using only one single-shot measurement per iteration. In comparison, if we want to calculate a given expectation value, we require at least $1000$ single-shot measurements for a single qubit. Then for this case, our algorithm requires less resources than any other classical-quantum algorithm that utilizes expectation values. For the VQE algorithm, first we choose $500$ single-shot measurements per step and COBYLA as the classical optimization method.  VQE needs $33$ COBYLA iterations to converge, which means $16500$ single-shot measurements in total, \ie 100 times the resources needed in our algorithm, and get a fidelity of $0.997$. If we change the number of single-shot measurements to $8192$ per step (it is the maximum shots allowed by IBM), we need $35$ COBYLA iterations to converge, which means $286720$ single-shot measurements, $1000$ times more resources than our algorithms, nevertheless, the fidelity is $0.999$.

\begin{equation}
2.\quad\tau\hat{\mathcal{O}}=\frac{\pi}{4}\sigma_x\Rightarrow\Delta=\frac{\pi}{2}\Rightarrow\mathcal{F}=\frac{1}{2}(1+\sqrt{2P_0-1}).\nonumber
\end{equation}
Now, we choose the reward ratio $r=0.9$ and the punishment ratio $p=1.5/r$. The results of the $40$ experiments are collected in the Appendix Table 2 (see supplemental material) and summarized in the histograms of Fig. \ref{Fig04}. From figure \ref{Fig04} (a) we can see that the probability $P_0$ is bigger than 0.9 in $35$ cases, which implies, as is shown in Fig. \ref{Fig04} (b), that most cases give fidelities larger than $0.94$. Also, we have $30$ experiments with $\mathcal{F}> 0.96$, the average fidelity is $\bar{\mathcal{F}}=0.97$ and the standard deviation is $\sigma=0.022$ which represent the $2.3\%$ of the average fidelity $\bar{\mathcal{F}}$. Also, the average number of iterations of the algorithm in the $40$ experiments is $\bar{N}=116$, the minimum number of iterations $N_{min}=25$ and the maximum number of iterations $N_{max}=572$, again for this case our algorithm uses less resources than the algorithm that use expectation values. As in the previous case, we compare the results with the VQE algorithm. For $500$ shots per step, we get a fidelity of $0.883$ with $23$ COBYLA iterations, which means $11500$ single-shot measurements, \ie $100$ times more resources than our algorithm. For $8192$ shots per step, the fidelity is $0.891$ and we need $23$ COBYLA iterations, the total single-shot measurements are $188416$, \ie 1000 times more resources than in our algorithm.

\begin{figure}[t]
\centering
\includegraphics[width=0.6\linewidth]{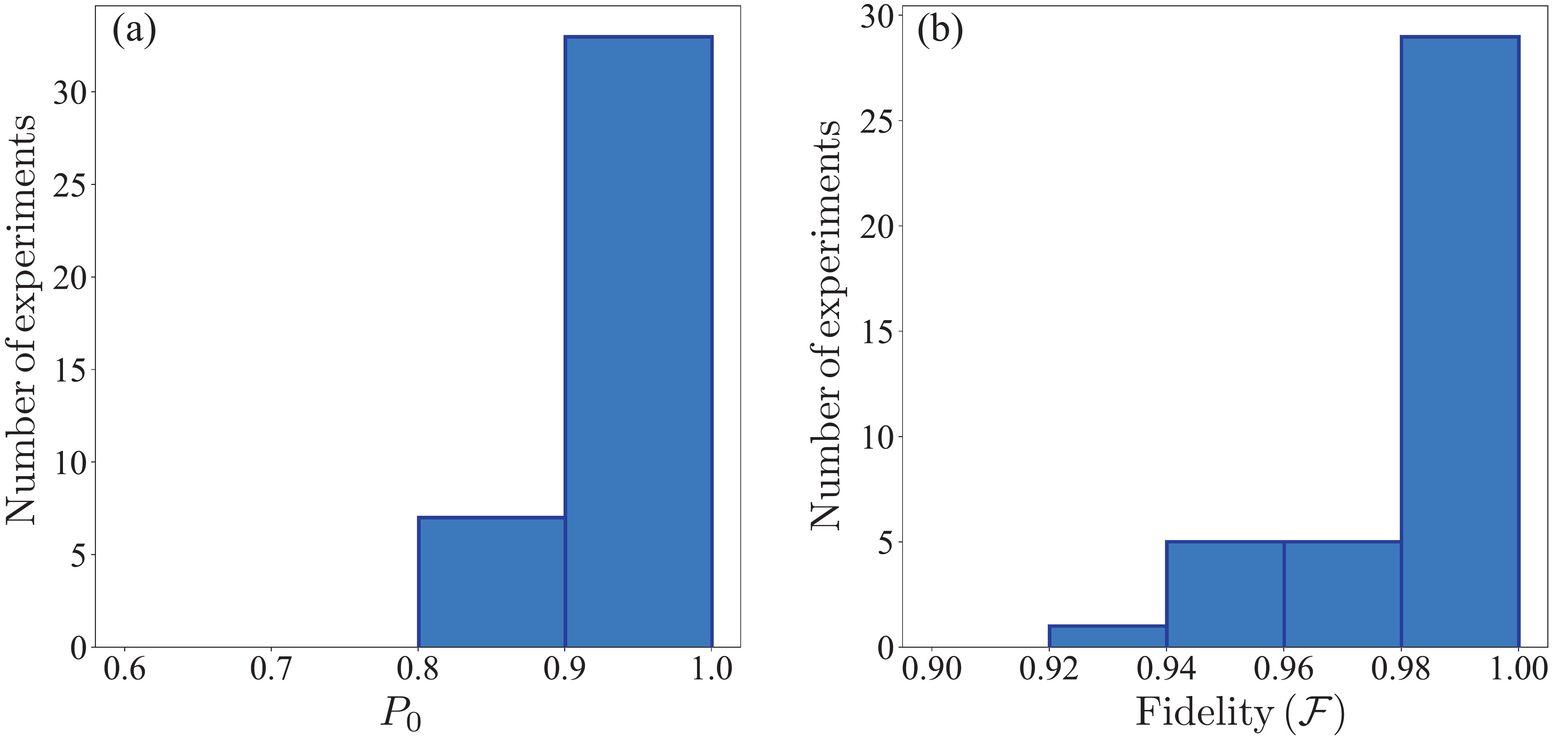}
\caption{Histograms for the results of $40$ independent experiments. with $\tau\hat{\mathcal{O}}=\cos{\frac{1}{10}}\sigma_x+\sin{\frac{1}{10}}\sigma_y$, $r=0.9$ and $p=1.5/r$. (a): Histogram for the probability to obtain $m=0$. (b): Histogram for the fidelity between the agent and the nearest eigenvector using Eq.~(\ref{Eq17}).}
\label{Fig05}
\end{figure}

\begin{eqnarray}
3.\quad&&\tau\hat{\mathcal{O}}=\cos{\frac{1}{10}}\sigma_x+\sin{\frac{1}{10}}\sigma_y\Rightarrow\Delta=2\nonumber\\
&&\Rightarrow\mathcal{F}=\frac{1}{2}\left(1+\sqrt{1+\frac{2(P_0-1)}{1-\cos{2}}}\right)\nonumber
\end{eqnarray}
We choose the reward ratio $r=0.9$ and the punishment ratio $p=1.5/r$ as in the previous case. The results of the $40$ experiments are collected in the Appendix Table 3 (see supplemental material) and summarized in the histograms of Fig. \ref{Fig05}. From Fig.~\ref{Fig05} (a) we can see that the probability $P_0$ is bigger than 0.85 in $39$ cases, which implies, as is shown in Fig. \ref{Fig05} (b), that most cases give fidelities larger than $0.94$. Also, we have $30$ experiments with $\mathcal{F}> 0.98$, the average fidelity is $\bar{\mathcal{F}}=0.98$ and the standard deviation of $\sigma=0.015$ which represent the $1.6\%$ of the average fidelity $\bar{\mathcal{F}}$. Also, the average number of iterations of the algorithm in the $40$ experiments was $\bar{N}=227$, the minimum number of iterations $N_{min}=26$ and the maximum number of iterations $N_{max}=782$. In this case, as $N_{max}$ is around 800, we compare the VQE algorithm, at first with $800$ shots per step, obtaining a fidelity of $0.911$ using $14$ COBYLA iterations, which means, a total number of single-shot measurements of $11200$, \ie 50 times more resources than our algorithms. When we use  $8192$ per step, the fidelity is $0.999$ and we need $14$ COBYLA iterations, obtaining a total number of single-shot measurements of $114688$, \ie $500$ times more resources than our algorithm.

Even if VQE allows us to reach fidelities larger than $0.98$ (the mean fidelity of our algorithm), it needs several resources, more than 100 times the resources using by our algorithm, which implies a great advantage of our proposal.

\subsection{Two-qubit case}
In this case, we have three different agent states given by
\begin{eqnarray}
&&\ket{\mathcal{A}_k^{(0)}}=\hat{D}_k\ket{00}, \nonumber\\
&&\ket{\mathcal{A}_k^{(1)}}=\hat{D}_k\ket{01}, \nonumber\\
&&\ket{\mathcal{A}_k^{(2)}}=\hat{D}_k\ket{10}.
\label{Eq18}
\end{eqnarray}
We update the matrix $\hat{D}_k$ according to Eq. (\ref{Eq12}). To decompose the matrix $\hat{D}_k$ in a set of one- and two-qubit gates, we use the method already implemented in qiskit~\cite{ComandWeb}. To find all the eigenvectors we divide the protocol in three stages. In the first stage, we consider the agent state $\ket{\mathcal{A}_k^{(0)}}=\hat{D}_k\ket{00}$, with $\hat{D}_1=\mathbb{I} $ and $w_1=1$. The outcome of the measure have four possibilities $m\in\{00,\,01,\,10,\,11\}$ and we run the algorithm until $w_{n_1}<0.1$ ($n_1$ iterations). After this, we have that $\ket{A_{n_1}^{(0)}}=\hat{D}_{n_1}\ket{00}$ is the approximation of one of the eigenvectors of $\hat{\mathcal{O}}$. 

In the second stage, we consider the agent state $\ket{\mathcal{A}_k^{(1)}}=\hat{D}_k\ket{01}$, with $\hat{D}_{n_1+1}=\hat{D}_{n_1} $ and $w_{n_1+1}=1$. Now, we take into account only three outcome $m\in\{01,\,10,\,11\}$, since we suppose that $\ket{\mathcal{A}_{N_1}^{(0)}}$ is a good enough approximation. If we obtain $m=00$, we consider it as an error, and we define $\hat{D}_{k+1}=\hat{D}_k$ and $w_{k+1}=w_k$, it means that we do nothing, and not apply the updating rule for $\hat{D}_{k+1}$ and $w_{k+1}$, we denote this error as $c_{00}$. We run this stage $n_2$ iterations until $w_{n_1+n_2}<0.1$. As we do not do rotations in the subspace spanned by $\{\ket{00},\,\ket{01}\}$ during this stage, we have $\ket{\mathcal{A}_{n_1+n_2}^{(0)}}=\ket{\mathcal{A}_{n_1}^{(0)}}$. Now, we obtain the approximation of two eigenvectors $\ket{A_{n_1+n_2}^{(1)}}=\hat{D}_{n_1+n_2}\ket{01}$ and $\ket{A_{n_1+n_2}^{(0)}}=\hat{D}_{n_1+n_2}\ket{00}$.

Finally, in the third stage, we consider the agent state $\ket{\mathcal{A}_k^{(2)}}=\hat{D}_k\ket{10}$, with $\hat{D}_{n_1+n_2+1}=\hat{D}_{n_1+n_2} $ and $w_{n_1+n_2+1}=1$. Now, we have only two possibilities for the outcome measurement $m\in\{10,\,11\}$. Here, we also suppose that $\hat{D}_{n_1+n_2}\ket{00}$ and $\hat{D}_{n_1+n_2}\ket{01}$ are good enough approximations. If we obtain $m=00$ or $m=01$, we consider them again as an error and we do not apply the update rule, denoting these errors as $c_{00}^{'}$ and $c_{01}$, like in the previous stage. We run this case $n_3$ iterations until $w_{n_1+n_2+n_3}<0.1$. In this stage, we only modify the subspace expanded by $\{\ket{10},\,\ket{11}\}$, then, we have that $\ket{\mathcal{A}_{n_1+n_2+n_3}^{(0)}}=\ket{\mathcal{A}_{n_1+n_2}^{(0)}}=\ket{\mathcal{A}_{n_1}^{(0)}}$ and $\ket{\mathcal{A}_{n_1+n_2+n_3}^{(1)}}=\ket{\mathcal{A}_{n_1+n_2}^{(1)}}$. After this procedure we obtained the approximation of all the eigenvectors $\{\ket{A_{n_T}^{(0)}}=\hat{D}_{n_T}\ket{00},\,\ket{A_{n_T}^{(1)}}=\hat{D}_{n_T}\ket{01},\,\ket{A_{n_T}^{(2)}}=\hat{D}_{n_T}\ket{10},\,\ket{A_{n_T}^{(3)}}=\hat{D}_{n_T}\ket{11}\}$, with $n_T=n_1+n_2+n_3$.

To test the algorithm, we choose three cases. First we consider the bi-local operator given by

\begin{equation}
1. \quad\tau\hat{\mathcal{O}}=\sigma_x\sigma_x=
\begin{pmatrix} 0 & 0& 0 & 1\\
			0 & 0 & 1 & 0\\
			0 & 1 & 0 & 0\\
			1 & 0 & 0 & 0
\end{pmatrix}.
\label{Eq19}
\end{equation}

In this case,  the eigenstates and the eigenvalues are
\begin{eqnarray}
	\ket{\mathcal{E}^{(0)}}=&&\frac{1}{\sqrt{2}}(\ket{00}-\ket{11}), \quad \alpha^{(0)}=-1, \nonumber\\
	\ket{\mathcal{E}^{(1)}}=&&\frac{1}{\sqrt{2}}(-\ket{01} + \ket{10}), \quad \alpha^{(1)}=-1, \nonumber\\
	\ket{\mathcal{E}^{(2)}}=&&\frac{1}{\sqrt{2}}(\ket{00}+\ket{11}), \quad \alpha^{(2)}=1 ,\nonumber\\
	\ket{\mathcal{E}^{(3)}}=&&\frac{1}{\sqrt{2}}(\ket{01}+\ket{10}), \quad \alpha^{(3)}=1.
	\label{Eq20}
\end{eqnarray}

We note that the ground state is degenerate, then any linear state of the form $\ket{\phi}=a\ket{\mathcal{E}^{(0)}}+b\ket{\mathcal{E}^{(1)}}$ will be also ground state of the operator and the same for the other states. In this case we define the fidelity of our algorithm by the probability to measure the initial state $\ket{j}$
\begin{equation}
	\mathcal{F}_j=P_j=|\langle j | \hat{D}^{\dagger}_{n_T}\hat{E}\hat{D}_{n_T}| j \rangle |^{2}.
\end{equation}
We run this case using IBM backend ``ibmq\_vigo"  and the results are shown in Appendix Table 4 (see supplemental material).
In this case, we run the algorithm ten times and the mean fidelities are: $\mathcal{F}_{00}=0.931$, $\mathcal{F}_{01}=0.933$, $\mathcal{F}_{10}=0.932$, and $\mathcal{F}_{11}=0.919$. The mean number of iterations is $\bar{N}=272$. In this case, the mean errors are: $\bar{c}_{00}=10$, $\bar{c}_{00}^{'}=8$ and $\bar{c}_{01}=5$. Therefore, the fidelity of our algorithm was higher than  $0.91$ for each eigenstate in less than $300$ single-shot measurements. The same as the single-qubit case, we will compare with the $VQE$ algorithm. At first, we choose $300$ shots per step, and $56$ COBYLA iterations, which means $16800$ single-shot measurements, obtaining a fidelity of $0.976$ for the ground state. Using  $8192$ shots per step, VQE needs $54$ COBYLA iterations to converge, which means $442368$  single-shot measurements, obtaining a fidelity of $0.997$ for the ground state. In this case, VQE get a significantly more accurate result, but it is only for the ground state and uses $1000$ times more resources than our algorithm which obtain all the eigenvectors.

The second example is the molecular hydrogen Hamiltonian with a bound length of $0.2~[\AA]$ \cite{Malley2016}:
\begin{equation}
	H=g_0\mathbb{I}+g_1Z_0+g_2Z_1+g_3Z_0Z_1+g_4Y_0Y_1+g_5X_0X_1,
\end{equation}
with $g_0=2.8489, g_1=0.5678, g_2=-1.4508, g_3=0.6799, g_4=0.0791, g_5=0.0791$. In this case the environment is given by

\begin{equation}
2.\quad \tau\hat{\mathcal{O}}=\begin{pmatrix}
		g_0+g_1+g_2+g_3 & 0 & 0& g_5-g_4\\
		0 & g_0+g_1-g_2-g_3 & g_4+g_5 & 0\\
		0 & g_4+g_5 & g_0-g_1+g_2-g_3 & 0\\
		g_5-g_4 & 0 & 0 & g_0-g_1-g_2+g_3
						\end{pmatrix},
\end{equation}
with the next eigenvectors and eigenvalues
\begin{eqnarray}
	\ket{\mathcal{E}^{(0)}}=&&-0.03909568\ket{01}+0.99923547\ket{10}, \quad \alpha^{(0)}=0.14421033,\nonumber\\
	\ket{\mathcal{E}^{(1)}}=&&\ket{00}, \quad \alpha^{(1)}=2.6458, \nonumber\\
	\ket{\mathcal{E}^{(2)}}=&&0.99923547\ket{01}+0.03909568\ket{10}, \quad \alpha^{(2)}=4.19378967,\nonumber\\
	\ket{\mathcal{E}^{(3)}}=&&\ket{11}, \quad \alpha^{(3)}=4.4118
\end{eqnarray}

In this case, we choose the same method as the previous case to calculate the $\mathcal{F}$, we choose IBM backend ``ibmq\_valencia" and the results are shown in Appendix Table 5 (see supplemental material). In this case, we run the algorithm ten times and the mean fidelities are: $\mathcal{F}_{00}=0.989$, $\mathcal{F}_{01}=0.973$, $\mathcal{F}_{10}=0.976$ and $\mathcal{F}_{11}=0.979$. The mean errors are: $\bar{c}_{00}=7$, $\bar{c}_{00}^{'}=4$ and $\bar{c}_{01}=3$ and the mean number of iterations is $\bar{N}=111$. In this case, we need less than $150$ single-shot measurements to obtain the fidelity over $0.97$. For the VQE algorithm, at first we choose $120$ shots per step and we need to use $59$ COBYLA iterations, which means $7080$ single-shot measurements, obtaining a fidelity of $0.994$ for the ground state. When we use $8192$ shots per step and VQE needs $64$ COBYLA iterations to converge, it means $507904$ single-shot measurements, obtaining a fidelity of $0.999$ for the ground state. In this case, VQE can get better fidelities (larger than 0.99) but use again much more resources than our proposal, around  $1000$ times more to get only one of the eigenvectors.

The third case that we consider to test the algorithm is the non-degenerate two-qubit operator

\begin{equation}
3. \quad	\tau\hat{\mathcal{O}}=\begin{pmatrix}
		\pi & -\frac{\pi}{2} & -\frac{\pi}{4} & -\frac{\pi}{4} \\
		-\frac{\pi}{2} & \pi & -\frac{\pi}{4} & -\frac{\pi}{4} \\
		-\frac{\pi}{4} & -\frac{\pi}{4} & \frac{\pi}{2} & 0\\
		-\frac{\pi}{4} & -\frac{\pi}{4} & 0 & \frac{\pi}{2} 
	\end{pmatrix},
	\label{Eq21}
\end{equation}
with eigenvectors and eigenvalues given by
\begin{eqnarray}
	\ket{\mathcal{E}^{(0)}}=&&\frac{1}{2}(\ket{00}+\ket{01}+\ket{10}+\ket{11}), \quad \alpha^{(0)}=0, \nonumber\\
	\ket{\mathcal{E}^{(1)}}=&&\frac{1}{\sqrt{2}}(\ket{10} - \ket{11}), \quad \alpha^{(1)}=\frac{\pi}{2}, \nonumber\\
	\ket{\mathcal{E}^{(2)}}=&&\frac{1}{2}(\ket{00}+\ket{01}-\ket{10}-\ket{11}), \quad \alpha^{(2)}=\pi ,\nonumber\\
	\ket{\mathcal{E}^{(3)}}=&&\frac{1}{\sqrt{2}}(\ket{00}-\ket{01}), \quad \alpha^{(3)}=\frac{3\pi}{2}.
	\label{Eq22}
\end{eqnarray}

We run the algorithm in the IBM quantum computer ``ibmq\_vigo". In order to reduce the total number of iterations, we run the three stages of the algorithm four times as follows:
\begin{enumerate}
\item We choose $r=0.6,\,p=1/r,\,\hat{D}_1=\mathbb{I},\, w_1=1$. Suppose that the total number of iteration after the three stages is $N_1=\eta_1$.
\item We choose $r=0.7,\,p=1/r,\,\hat{D}_{\eta_1+1}=\hat{D}_{\eta_1},\, w_{\eta_1+1}=1$. Suppose that the total number of iteration after the three stages is $N_2=\eta_1+\eta_2$.
\item We choose $r=0.8,\,p=1/r,\,\hat{D}_{N_2+1}=\hat{D}_{N_2},\, w_{N_2+1}=1$. Suppose that the total number of iteration after the three stages is $N_3=\eta_1+\eta_2+\eta_3$.
\item We choose $r=0.9,\,p=1/r,\,\hat{D}_{N_3+1}=\hat{D}_{N_3},\, w_{N_3+1}=1$, and suppose that the total number of iteration after the three stages is $N=\eta_1+\eta_2+\eta_3+\eta_4$.
\end{enumerate}
We define the fidelity of each approximation as

\begin{equation}
	\mathcal{F}_{\ell m}=\max \limits_{k=\{0,1,2,3\}} |\bra{\mathcal{E}^{(k)}}\hat{D}_{N}\ket{\ell m}|^2.
	\label{Eq23}
\end{equation}

To obtain a data set to evaluate the performance of our protocol, we perform ten independent experiments. These data are collected in Appendix Table 6 (see supplemental material). The average fidelities that we obtain are $\bar{\mathcal{F}}_{00}=0.941,\,\bar{\mathcal{F}}_{01}=0.933,\,\bar{\mathcal{F}}_{10}=0.929,\,\bar{\mathcal{F}}_{11}=0.935$, the average number of iterations is $\bar{N}=1396$ and the mean errors are: $\bar{c}_{00}=29$, $\bar{c}_{00}^{'}=19$ and $\bar{c}_{01}=18$. Therefore, in this case we obtain the four eigenvectors with fidelities larger than $0.92$ in less than $1500$ single-shot measurements, which at least corresponds to $6$ measurements of mean values, being not enough for a classical-quantum algorithm that uses the optimization of mean values. For the VQE algorithm, we choose $2000$ shots per step using $77$ COBYLA iterations, which means $157000$ single-shot measurements obtaining a fidelity of $0.918$ for the ground state. For $8192$ shots per step, VQE needs $88$ COBYLA iterations to converge, it means  $720896$ single-shot measurements obtaining a fidelity of $0.944$. In this case, VQE cannot surpass the performance of our algorithm, and use more than $100$ times resources than our proposal only for the ground state.

For $n-$qubit observable ($n>2$), we can use the same protocol but considering more measurement outputs, which implies more stages in the algorithm.

\section{Conclusions}
In this work, we implement satisfactorily the approximate eigensolver~\cite{AlbarranArriagada2020} using the IBM quantum computer. For the single-qubit case, we obtain fidelities larger than $0.97$ for both eigenvectors using around $200$ single-shot measurements. For the two-qubit case, we use around $1500$ single-shot measurements to obtain the approximation of the four eigenvectors with fidelity over $0.9$. Due to the stochastic nature of this protocol, we cannot ensure that the approximation converges asymptotically with the number of iteration to the eigenvectors. Nevertheless, it is useful to obtain a fast approximation to use as a guess into another eigensolver that can reach maximal fidelity, like in the eigensolver of Ref.~\cite{Wei2020}. Also, we compare the performance of our proposal with the VQE algorithm, where VQE, in general, get better fidelities in the single-qubit case but use more than 100 times the number of resources than our algorithm. For two-qubit, the advantage in the maximal fidelity of VQE is a little better in comparison with our algorithm, but again, VQE needs several resources, \ie more than 1000 times the resources used by our algorithm for all the eigenvectors. Also, the performance of the VQE algorithm depends on the variational ansatz used, which is not the case with our algorithm. This dependence of the VQE algorithms allows enhancing its performance using a better ansatz. The main goal of our algorithm is to get a high fidelity approximation for all the eigenvectors with few resources. This goal is completely satisfied in comparison with the resources needed for VQE. On the other hand, by manipulating the convergence criteria of our algorithm, we can reach better fidelities. Finally, this work also paves the way for the development of future suitable quantum devices to work with limited resources.

\section*{Acknowledgments}
We acknowledge financial support from Spanish MCIU/AEI/FEDER (PGC2018-095113-B-I00), Basque Government IT986-16, projects QMiCS (820505) and OpenSuperQ (820363) of EU Flagship on Quantum Technologies, EU FET Open Grant Quromorphic, EPIQUS, and Shanghai STCSM (Grant No. 2019SHZDZX01-ZX04).

\section*{Author Contributions} E.S. and F.A.-A. supervised and contributed to the theoretical analysis. N.B. carried out all calculations and prepared the figures. C.-Y.P. and M.H write the qiskit program to run in IBM quantum experience. All the authors wrote the manuscript. All authors contributed to the results discussion and revised the manuscript.

\textbf{Competing Interests:} The authors declare that they have no competing interests.

\textbf{Data Availability:} The qiskit codes of the one-qubit case and the two-qubit case are available in \href{https://github.com/Panchiyue/Qiskit-Code/tree/main}{https://github.com/Panchiyue/Qiskit-Code/tree/main}.

\clearpage

{\Huge{Supplemental material}}
\begin{appendix} 
\captionsetup[table]{labelfont={bf},labelformat={default},labelsep=period,name={Appendix Table}}

\section{Derivation of Eq. (17).}
\label{AppA}
Using Eq. (2) and Eq. (3) of the main text, the operator $\hat{\mathcal{O}}\tau$ for a two-level system is write as
\begin{eqnarray}
	\hat{\mathcal{O}}=\alpha^{(0)}\ket{\mathcal{E}^{(0)}}\bra{\mathcal{E}^{(0)}}+\alpha^{(1)}\ket{\mathcal{E}^{(1)}}\bra{\mathcal{E}^{(1)}},
	\label{EqA01}
\end{eqnarray}
then, the environment operator E given by 
\begin{equation}
\hat{E}=e^{-i\hat{\mathcal{O}}\tau}=e^{-i\alpha^{(0)}}\ket{\mathcal{E}^{(0)}}\bra{\mathcal{E}^{(0)}}+e^{-i\alpha^{(1)}}\ket{\mathcal{E}^{(1)}}\bra{\mathcal{E}^{(1)}},
\label{EqA02}
\end{equation}

According to Eq. (14) (main text),  the agent state at the end of the protocol (after $N$ iterations) reads
\begin{equation}
	\ket{\mathcal{A}_N^{(j)}}=\hat{D}_N\ket{j},
\end{equation}
with $j=\{0,1\}$, it means 
\begin{equation}
	\hat{D}_N=\ket{\mathcal{A}_N^{(0)}}\bra{0}+\ket{\mathcal{A}_N^{(1)}}\bra{1}.
\end{equation}

Without loss of generality, we suppose that the fidelity $\mathcal{F}=|\braket{\mathcal{E}^{(0)}}{\mathcal{A}_N^{(0)}}|^2 > |\braket{\mathcal{E}^{(0)}}{\mathcal{A}_N^{(1)}}|^2$, then
\begin{equation}
	\ket{\mathcal{A}_N^{(0)}}=\sqrt{\mathcal{F}}\ket{\mathcal{E}^{(0)}}+e^{i\varphi}\sqrt{1-\mathcal{F}}\ket{\mathcal{E}^{(0)}},
\end{equation}
$\varphi\in[0,2\pi]$. As $\ket{\mathcal{A}_N^{(0)}}$ and $\ket{\mathcal{A}_N^{(1)}}$ are orthogonal, we have
\begin{equation}
	\ket{\mathcal{A}_N^{(1)}}=\sqrt{1-\mathcal{F}}\ket{\mathcal{E}^{(0)}}-e^{i\varphi}\sqrt{\mathcal{F}}\ket{\mathcal{E}^{(0)}}.
\end{equation}

Now, the probability to measure the state $\ket{0}$ at the end of the protocol (after $N$ iterations is given by
\begin{eqnarray}
	&&P_0=|\langle 0 |\hat{D}^{\dagger}_N E \hat{D}_N | 0 \rangle |^2\nonumber\\
	&&=|\langle 0 |\left(\ket{0}\bra{\mathcal{A}_N^{(0)}}+\ket{1}\bra{\mathcal{A}_N^{(1)}}\right) \left(e^{-i\alpha^{(0)}} \ket{\mathcal{E}^{(0)}}\bra{\mathcal{E}^{(0)}}+e^{-i\alpha^{(1)}} \ket{\mathcal{E}^{(1)}}\bra{\mathcal{E}^{(1)}}\right) \left(\ket{\mathcal{A}_N^{(0)}}\bra{0}+\ket{\mathcal{A}_N^{(1)}}\bra{1}\right) | 0 \rangle |^2\nonumber\\
	&&=|\bra{\mathcal{A}_N^{(0)}}\left(e^{-i\alpha^{(0)}} \ket{\mathcal{E}^{(0)}}\bra{\mathcal{E}^{(0)}}+e^{-i\alpha^{(1)}} \ket{\mathcal{E}^{(1)}}\bra{\mathcal{E}^{(1)}}\right)\ket{\mathcal{A}_N^{(0)}}|^2=|e^{-i\alpha^{(0)}}\mathcal{F}+e^{-i\alpha^{(1)}}(1-\mathcal{F})|^2\nonumber\\
	&&=\left[e^{-i\alpha^{(0)}}\mathcal{F}+e^{-i\alpha^{(1)}}(1-\mathcal{F})\right]\left[e^{i\alpha^{(0)}}\mathcal{F}+e^{i\alpha^{(1)}}(1-\mathcal{F})\right]=\mathcal{F}^2+\left(e^{i\Delta}+e^{-i\Delta}\right)\mathcal{F}(1-\mathcal{F})+(1-\mathcal{F})^2\nonumber\\
	&&=\mathcal{F}^2 + [\mathcal{F}^2-2\mathcal{F}+1]+2\cos(\Delta)\mathcal{F}(1-\mathcal{F})\nonumber\\
	&&\Rightarrow P_0=2\mathcal{F}(\mathcal{F}-1)\left[1-\cos(\Delta)\right] + 1,
\end{eqnarray}
with $\Delta=|\alpha^{(1)}-\alpha^{(0)}|$, recovering the expression given by Eq. (17) in the main text.

\clearpage
\section{DATA SETS OF SINGLE-QUBIT CASES}
\label{AppB}

\begin{table}[h]
\caption{Data set of $\hat{\mathcal{O}\tau}=\frac{\pi}{2}\sigma_x$}  
\begin{tabular*} {15cm}{llllllllllll}  
\hline  
\hline
$\textrm{Ex}\footnotemark[1]\footnotetext[1]{Experiment number}$ & 1 & 2 & 3 & 4 & 5 & 6 & 7 & 8 & 9 & 10\\ 
\hline 
$N$ & 51 & 59 & 52 & 167 & 112 & 205 & 54 &116 & 57 & 43 \\  
$P_0$ & 0.981 & 0.963 & 0.884 & 0.980 & 0.947 & 0.990 & 0.969 & 0.706 & 0.895 & 0.940 \\ 
$\mathcal{F}$ & 0.995 & 0.991 & 0.970 & 0.995 & 0.987 & 0.997 & 0.992 & 0.920 & 0.973 & 0.985\\ 
\hline 
$\textrm{Ex}$ &11 & 12 & 13 & 14 & 15 & 16 & 17 & 18 & 19 & 20\\  
\hline  
$N$ &  185 & 162 & 107 & 113 & 64 & 64 & 96 & 190 & 42 & 111 \\  
$P_0$ & 0.893 & 0.928 & 0.782 & 0.972 & 0.836 & 0.917 & 0.683 & 0.996 & 0.983 & 0.981\\  
$\mathcal{F}$ & 0.972 & 0.982 & 0.942 & 0.993 & 0.957 & 0.978 & 0.913 & 0.991 & 0.996 & 0.995\\
\hline  
$\textrm{Ex}$ & 21 & 22 & 23 & 24 & 25 & 26 & 27 & 28 & 29 & 30\\  
\hline  
$N$ &  32 & 79 & 61 & 25 & 161 & 86 & 107 & 32 & 28 & 528  \\  
$P_0$ & 0.996 & 0.896 & 0.977 & 0.974 & 0.950 & 0.913 & 0.984 & 0.977 & 0.982 & 0.946 \\  
$\mathcal{F}$ & 0.991& 0.973 & 0.994 & 0.993 & 0.987 & 0.978 & 0.996 & 0.994 & 0.995 & 0.986\\
\hline  
$\textrm{Ex}$ & 31 & 32 & 33 & 34 & 35 & 36 & 37 & 38 & 39 & 40\\  
\hline  
$N$ &  44 & 85 & 94 & 39 & 149 & 25 & 33 & 63 & 197 & 198\\  
$P_0$ & 0.858 & 0.854 & 0.919 & 0.970 & 0.889 & 0.930 & 0.978 & 0.889 & 0.936 & 0.949 \\  
$\mathcal{F}$ & 0.963& 0.962 & 0.979 & 0.992 & 0.971 & 0.982 & 0.994 & 0.971 & 0.984 & 0.987\\
\hline  
\hline 
\end{tabular*}  
\label{Tab01}
\end{table}

\begin{table}[h]
\caption{Data set of $\hat{\mathcal{O}}\tau=\frac{\pi}{4}\sigma_x$}  
\begin{tabular*} {15cm}{llllllllllll}  
\hline  
\hline 
$\textrm{Ex}$ & 1 & 2 & 3 & 4 & 5 & 6 & 7 & 8 & 9 & 10\\  
\hline  
$N$ & 48 & 55 & 36 & 287 & 28 & 55 & 348 & 572 & 78 & 284 \\  
$P_0$ & 0.930 & 0.981 & 0.910 & 0.850 & 0.940 & 0.952 & 0.820 & 0.936 & 0.901 & 0.960\\  
$\mathcal{F}$ & 0.964 & 0.990 & 0.953 & 0.918 & 0.969 & 0.976 & 0.900 & 0.967 & 0.948 & 0.980\\
\hline  
$\textrm{Ex}$ &11 & 12 & 13 & 14 & 15 & 16 & 17 & 18 & 19 & 20\\  
\hline  
$N$ &  93 & 45 & 26 & 92 & 34 & 34 & 25 & 37 & 55 & 46\\  
$P_0$ & 0.941 & 0.992 & 0.967 & 0.950 & 0.975 & 0.900 & 0.936 & 0.912 & 0.945 & 0.920\\  
$\mathcal{F}$ & 0.970 & 0.996 & 0.983 & 0.974 & 0.987 & 0.947 & 0.967 & 0.954 & 0.972 & 0.958 \\
\hline  
$\textrm{Ex}$ & 21 & 22 & 23 & 24 & 25 & 26 & 27 & 28 & 29 & 30\\  
\hline  
$N$ &  47 & 65 & 108 & 109 & 74 & 225 & 141 & 153 & 35 & 54\\  
$P_0$ & 0.850 & 0.878 & 0.952 & 0.987 & 0.980 & 0.943 & 0.990 & 0.985 & 0.962 & 0.953 \\  
$\mathcal{F}$ & 0.918 & 0.935 & 0.976 & 0.993 & 0.990 & 0.971 & 0.995 & 0.992& 0.980 & 0.976 \\
\hline  
$\textrm{Ex}$ & 31 & 32 & 33 & 34 & 35 & 36 & 37 & 38 & 39 & 40\\  
\hline  
$N$ &  114 & 152 & 163 & 125 & 112 & 287 & 55 & 185 & 55 & 108 \\  
$P_0$ & 0.963 & 0.945 & 0.935 & 0.960 & 0.975 & 0.890 & 0.982 & 0.958 & 0.962 &0.972  \\  
$\mathcal{F}$ & 0.981 & 0.972 & 0.966 & 0.979 & 0.987 & 0.941 & 0.991 & 0.979 & 0.980 & 0.986\\
\hline  
\hline 
\end{tabular*}  
\label{Tab02}
\end{table}  

\begin{table}[h]
\caption{Data set of $\hat{\mathcal{O}}\tau=\cos\frac{1}{10}\sigma_x+\sin\frac{1}{10}\sigma_y$}  
\begin{tabular*} {15cm}{llllllllllll}  
\hline  
\hline 
$\textrm{Ex}$ & 1 & 2 & 3 & 4 & 5 & 6 & 7 & 8 & 9 & 10\\  
\hline  
$N$ & 55 & 49 & 26 & 138 & 320 & 95 & 98 & 31 & 287 & 170 \\  
$P_0$ & 0.956 & 0.945 & 0.98 & 0.916 & 0.889 & 0.951 & 0.868 & 0.976 & 0.989 & 0.989\\  
$\mathcal{F}$ & 0.984 & 0.980 & 0.993 & 0.969 & 0.959 & 0.982 & 0.951 & 0.991 & 0.996 & 0.996\\
\hline  
$\textrm{Ex}$ &11 & 12 & 13 & 14 & 15 & 16 & 17 & 18 & 19 & 20\\  
\hline  
$N$ &  341 & 221 & 156 & 196 & 180 & 255 & 782 & 186 & 496 & 183\\  
$P_0$ & 0.972 & 0.978 & 0.956 & 0.982 & 0.978 & 0.923 & 0.965 & 0.956 & 0.854 & 0.959\\  
$\mathcal{F}$ & 0.990 & 0.992 & 0.984 & 0.994 & 0.992 & 0.972 & 0.987 & 0.984 & 0.945 & 0.985 \\
\hline  
$\textrm{Ex}$ & 21 & 22 & 23 & 24 & 25 & 26 & 27 & 28 & 29 & 30\\  
\hline  
$N$ &  198 & 98 & 191 & 158 & 125 & 186 & 165 & 145 & 155 & 58\\  
$P_0$ & 0.955 & 0.895 & 0.994 & 0.965 & 0.948 & 0.856 & 0.962 & 0.952 & 0.966 & 0.952 \\  
$\mathcal{F}$ & 0.984 & 0.961 & 0.998 & 0.987 & 0.981 & 0.946 & 0.984 & 0.982 & 0.988 & 0.983 \\
\hline  
$\textrm{Ex}$ & 31 & 32 & 33 & 34 & 35 & 36 & 37 & 38 & 39 & 40\\  
\hline  
$N$ &  493 & 435 & 156 & 327 & 535 & 254 & 423 & 138 & 75 & 556 \\  
$P_0$ & 0.943 & 0.972 & 0.944 & 0.954 & 0.973 & 0.946 & 0.955 & 0.876 & 0.963 & 0.82  \\  
$\mathcal{F}$ & 0.979 & 0.990 & 0.978 & 0.983 & 0.990 & 0.980 & 0.984 & 0.954 & 0.987 & 0.932\\
\hline  
\hline 
\end{tabular*}  
\label{Tab03}
\end{table}

\clearpage
\section{DATA SET OF TWO-QUBIT CASE}
\label{AppC}

In this appendix, we will show the all results of two-qubit case, the first line ``EX" means that we had run this case 5 times, the second line ``$N$" is the total iterations for each time, the third line ``$c_{00}$" is the number of error ``00" when the input state is $\ket{01}$, the forth line  and the fifth line ``$c_{00}^{'}$" and ``$c_{01}$" are the times of error ``00" and ``01", respectively, when input state is $\ket{10}$. The finial four lines are the fidelities of the eigenstates. 

\begin{table}[h]
\centering
\caption{Data set for $\hat{\mathcal{O}}\tau$ given by Eq. (19)}  
\begin{tabular*} {13cm}{llllllllllll}  
\hline  
\hline
$\textrm{Ex}$ & 1 & 2 & 3 & 4 & 5 & 6 & 7 & 8 & 9 & 10\\ 
\hline
$N$ & 306 & 304 & 188 & 253 & 303 & 219 & 197 & 412 & 130 &410\\
\hline

$c_{00}$ & 19 & 15 & 10 & 14 & 11 & 12 & 3 & 7 & 2 & 3\\

\hline
$c_{00}^{'}$ & 31 & 0 & 6 & 7 & 4 & 1 & 1 & 10 & 5 & 16\\
\hline

$c_{01}$ & 11 & 3 & 8 & 2 & 1 & 0 & 5 &4 & 8 & 9\\
\hline
$\mathcal{F}_{00}$ &0.926  & 0.915 & 0.911 & 0.916 & 0.929 & 0.925 & 0.946 & 0.931 & 0.966 & 0.942\\

$\mathcal{F}_{01}$ & 0.911 & 0.900 & 0.932 & 0.954 & 0.951 & 0.933 & 0.94 & 0.898 & 0.928 & 0.978\\

$\mathcal{F}_{10}$ & 0.912 & 0.932 & 0.925 & 0.911 & 0.912 & 0.989 & 0.932 & 0.912 & 0.96 & 0.938\\

$\mathcal{F}_{11}$ & 0.902 & 0.912 & 0.909 & 0.900 & 0.913 & 0.981 & 0.955 & 0.885 & 0.934 &0.903\\ 
\hline 
\hline  

\end{tabular*}  
\label{Tab04}
\end{table}

\begin{table}[h]
\centering
\caption{Data set of $\hat{\mathcal{O}}\tau$ given by Eq. (23)}  
\begin{tabular*} {13cm}{lllllllllll}  
\hline  
\hline
$\textrm{Ex}$ & 1 & 2 & 3 & 4 & 5 & 6 & 7 & 8 & 9 & 10\\ 
\hline
$N$ & 75 & 128 & 90 & 86 & 233 & 92 & 149 & 92 & 92 & 73 \\
\hline

$c_{00}$ & 1 & 5 & 2 & 4 & 6 & 4 & 2 & 4 & 4 & 3 \\
\hline

$c_{00}^{'}$ & 2 & 1 & 4 & 2 & 1 & 1 & 8 & 1 & 1 & 0\\
\hline

$c_{01}$ & 2 & 4 & 1 & 1 & 1 & 0 & 7 & 0 & 0 & 1\\
\hline

$\mathcal{F}_{00}$ &0.992  & 0.998 & 0.986 & 0.986 & 0.992  & 0.994 & 0.999 & 0.964 & 0.991 & 0.991\\

$\mathcal{F}_{01}$ & 0.996 & 0.942 & 0.984 & 0.997 & 0.990 & 0.994 & 0.947 & 0.956 & 0.940 & 0.988\\

$\mathcal{F}_{10}$ & 0.997 & 0.989 & 0.969 & 0.996 & 0.989 & 0.989 & 0.957 & 0.976 & 0.918 & 0.975\\

$\mathcal{F}_{11}$ & 0.991 & 0.943 & 0.971 & 0.997 & 0.988 & 0.992 & 0.987 & 0.988 & 0.959 & 0.978\\ 
\hline 
\hline  

\end{tabular*}  
\label{Tab05}
\end{table}

\begin{table}[h]
\centering
\caption{Data set of $\hat{\mathcal{O}}\tau$ given by Eq. (25)}  
\begin{tabular*} {13cm}{lllllllllll}  
\hline  
\hline
$\textrm{Ex}$ & 1 & 2 & 3 & 4 & 5 & 6 & 7 & 8 & 9 & 10\\ 
\hline
$N$ & 2370 & 1068 & 1702 & 1559 & 1711 & 1360 & 2174 & 431 & 1129 & 454\\
\hline

$c_{00}$ & 52 & 13 & 18 & 68 & 29 & 15 & 45 & 4 & 32 & 13\\
\hline

$c_{00}^{'}$ & 26 & 20 & 1 & 47 & 5 & 17 & 30 & 3 & 28 & 10\\
\hline

$c_{01}$ & 23 & 8 & 21 & 37 & 10 & 15 & 25 & 3 & 29 & 13\\
\hline

$\mathcal{F}_{00}$ &0.924  & 0.936 & 0.943 & 0.953 & 0.977 & 0.915 & 0.908 & 0.971 & 0.911 &0.971\\

$\mathcal{F}_{01}$ & 0.941 & 0.982 & 0.901 & 0.906 & 0.928  & 0.928 &0.898 & 0.975 & 0.923 & 0.946\\

$\mathcal{F}_{10}$ & 0.961 & 0.964 & 0.898 & 0.926 & 0.937 & 0.889 & 0.912 & 0.989 & 0.902 & 0.910 \\

$\mathcal{F}_{11}$ & 0.953 & 0.933 & 0.886 & 0.938 & 0.942 & 0.929 & 0.905 & 0.990 & 0.968 & 0.909 \\ 
\hline 
\hline  

\end{tabular*}  
\label{Tab06}
\end{table}

\end{appendix}

\end{document}